\documentclass[sigconf]{acmart}
\AtBeginDocument{%
  \providecommand\BibTeX{{%
    \normalfont B\kern-0.5em{\scshape i\kern-0.25em b}\kern-0.8em\TeX}}}

\begin{document}

%%
%% The "title" command has an optional parameter,
%% allowing the author to define a "short title" to be used in page headers.
\title{Study of Non-Verbal Behavior in Conversational Agents}

%%
%% The "author" command and its associated commands are used to define
%% the authors and their affiliations.
%% Of note is the shared affiliation of the first two authors, and the
%% "authornote" and "authornotemark" commands
%% used to denote shared contribution to the research.
% \author{Camila Vicari Maccari}
% \authornote{Both authors contributed equally to this research.}
% \email{trovato@corporation.com}
% \orcid{1234-5678-9012}
% \author{G.K.M. Tobin}
% \authornotemark[1]
% \email{webmaster@marysville-ohio.com}
% \affiliation{%
%   \institution{Institute for Clarity in Documentation}
%   \streetaddress{P.O. Box 1212}
%   \city{Dublin}
%   \state{Ohio}
%   \country{USA}
%   \postcode{43017-6221}
% }

\author{Camila Vicari Maccari}
\affiliation{%
  \institution{Pontifícia Universidade Católica do Rio Grande do Sul}
  \streetaddress{Avenida Ipiranga, 6681}
  \city{Porto Alegre}
  \country{Brazil}}
\email{camila.maccari@edu.pucrs.br}

\author{Gustavo Galle de Melo}
\affiliation{%
  \institution{Pontifícia Universidade Católica do Rio Grande do Sul}
  \streetaddress{Avenida Ipiranga, 6681}
  \city{Porto Alegre}
  \country{Brazil}}
\email{gustavo.melo01@edu.pucrs.br}

\author{Paulo Ricardo Knob}
\affiliation{%
  \institution{Pontifícia Universidade Católica do Rio Grande do Sul}
  \streetaddress{Avenida Ipiranga, 6681}
  \city{Porto Alegre}
  \country{Brazil}}
\email{paulo.knob@edu.pucrs.br}

\author{Soraia Raupp Musse}
\affiliation{%
  \institution{Pontifícia Universidade Católica do Rio Grande do Sul}
  \streetaddress{Avenida Ipiranga, 6681}
  \city{Porto Alegre}
  \country{Brazil}}
\email{soraia.musse@pucrs.br}

%%
%% By default, the full list of authors will be used in the page
%% headers. Often, this list is too long, and will overlap
%% other information printed in the page headers. This command allows
%% the author to define a more concise list
%% of authors' names for this purpose.
%\renewcommand{\shortauthors}{Trovato and Tobin, et al.}

%%
%% The abstract is a short summary of the work to be presented in the
%% article.
\begin{abstract}
  This paper studies the non-verbal behavior of a conversational agent named Arthur. We propose the development of body movements for this agent, which interacts solely through voice commands, chat, and videos with facial animations. This research aims to analyze users' perceptions regarding the gestures performed by Arthur. This study was conducted with participants who agreed to interact directly or through video with the conversational agent. The main goal is to analyze whether including nonverbal movements alters users' perception so that they feel more comfortable watching the video or interacting in real-time.
\end{abstract}

%%
%% The code below is generated by the tool at http://dl.acm.org/ccs.cfm.
%% Please copy and paste the code instead of the example below.
%%
%PA: tem que arrumar esse, caso precise mais coisas
\begin{CCSXML}
<ccs2012>
   <concept>
       <concept_id>10010147.10010371.10010352</concept_id>
       <concept_desc>Computing methodologies~Animation</concept_desc>
       <concept_significance>500</concept_significance>
       </concept>
 </ccs2012>
\end{CCSXML}

\ccsdesc[500]{Computing methodologies~Animation}

%%
%% Keywords. The author(s) should pick words that accurately describe
%% the work being presented. Separate the keywords with commas.
\keywords{Conversational Agents, Body Movements, Users, Non-verbal Behavior, Interaction}

%% A "teaser" image appears between the author and affiliation
%% information and the body of the document, and typically spans the
%% page.
\begin{teaserfigure}
  \includegraphics[width=\textwidth]{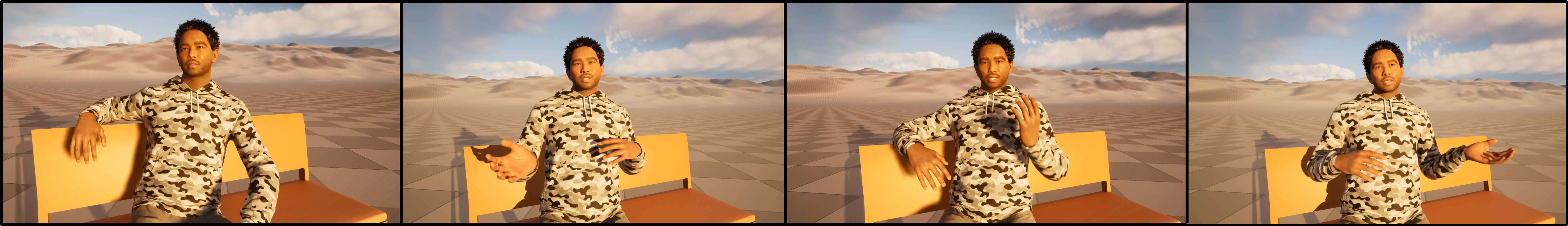}
  \caption{Illustrations of body animations designed for Arthur.}
  \Description{Illustrations of body animations designed for Arthur.}
  \label{fig:teaser}
\end{teaserfigure}

% \received{20 February 2007}
% \received[revised]{12 March 2009}
% \received[accepted]{5 June 2009}

%%
%% This command processes the author and affiliation and title
%% information and builds the first part of the formatted document.
\maketitle

\section{Introduction}

Conversation is an interactive dialogue that involves spoken language, facial expressions, and body gestures. These characteristics are intimately related, as they occur simultaneously and integrate the discourse of the communicators~\cite{cassell1994animated}. Therefore, developing conversational agents with these characteristics is essential as they increasingly approach human appearance. 

An Embodied Conversational Agent, or ECA, is an interactive character created on a computer. It possesses human appearance and behavior, interacting with users through dialogue, answering questions, and performing body movements~\cite{xiao2007role}. Additionally, conversational agents fulfill specific needs that conventional interfaces, such as mouse and keyboard, lack, making them an interesting choice for user interaction. They can engage in conversation, produce nonverbal behaviors, and provide feedback to the user during a dialogue~\cite{cassell2000embodied}. Therefore, more conversational agents with human-like behaviors are being created as they approach user characteristics. 

Virtual humans are becoming an interesting asset in human-computer interaction. Researchers are proposing methods to generate non-verbal behaviors, such as gestures and facial expressions~\cite{kucherenko2021genea}. First impressions are relevant, as they influence users' perception and willingness to interact with the agent again~\cite{biancardi2017could}. According to Nagy et al., ~\cite{nagy2021framework}, conversational agents are more appealing to users when they exhibit appropriate body movements. 

%PA: uma duvida aqui: o artigo do arthur, como citado aqui, é uma versão diferente do que foi usado nesse trabalho (arthur e bella naquele artigo, o "john" nesse; Unity naquele, Python nesse; Cartoon naquele, MH nesse; Animações naquele, vídeos nesse...). Pode dar problema? Talvez fosse bom deixar isso claro em algum lugar.
%SO: acho que tudo bem. São detalhes de implementação, a artqutetura de info é a mesma.
This paper aims to analyze the non-verbal behavior of a conversational agent named Arthur. This agent model was derived from the study by Knob et al.~\cite{knob2021arthur}, who developed a new conversational agent that communicates through voice commands, chat, and videos with facial animations. Therefore, we propose to develop body movements for this agent to assess users' opinions regarding the said body animations. By analyzing the results, it will be possible to evaluate the user experience and their perceptions of the proposed model. 

This work is organized as follows: Section~\ref{sec:relatedwork} presents the work related to our own. Section~\ref{sec:metodologia} describes the model developed for the conversational agent and how the experiment was conducted. Section~\ref{sec:resultados} presents the results obtained by this work. Finally, Section~\ref{sec:conclusao} presents the final considerations of this work and the possible future work.

\section{Related Work}
\label{sec:relatedwork}

This section presents some related works used as a basis for developing this project. This section will address studies with conversational agents that perform body movements and agents that interact with users through video or directly in real time. It is important to mention that we are not presenting technologies to provide facial animation e.g.,~\cite{Queiroz2008} and~\cite{Queiroz2010}, but methodologies used to provide higher level behaviors in virtual humans.

Pelachaud~\cite{pelachaud2005multimodal} proposed the creation of a conversational agent named Greta. The agent's behavior is synchronized with speech, consistent with pronouncing words and phrases. The verbal and non-verbal movements performed by Greta enable communication, allowing the agent to express himself/herself in various ways. The main goal of Pelachaud's work~\cite{pelachaud2005multimodal} is to study how emotions influence human-computer interaction. For an agent to initiate a conversation and maintain it with a user, it must be able to generate verbal and non-verbal behaviors and demonstrate emotions. According to the author, three computational domains are essential for conversational agents: perception, generation, and interaction, with the latter overlapping the other two. In perception, the Embodied Conversational Agent (ECA) must pay attention to the context in which it is inserted, such as perceiving the moment to initiate an interaction, how long to maintain the conversation, and when to give the turn back for the user to speak. In the interaction domain, the ECA has to interpret signals, emit, and analyze them. In a generation, the ECA must be able to reproduce visual and auditory behaviors in a synchronized and expressive manner. Among these domains, the paper mainly focuses on generation. To analyze the conversational agent, two studies were conducted with users, who were required to watch videos of the agent. Thus, the participant's perception of the different gestures performed by Greta was tested.

Breitfuss et al.~\cite{breitfuss2007automated} propose generating automatic verbal and non-verbal behaviors for virtual agents that engage in dialogue using a predefined dialogue script. Thus, the system controls both facial and bodily animations, as well as the speech of these agents. In the proposed model~\cite{breitfuss2007automated}, the behavior is generated for both the speaker and the listener. However, it is limited in that it only generates dialogues between the virtual agents themselves, so the user can only observe the dialogue and not participate in the interaction. 

According to Nagy et al., ~\cite{nagy2021framework}, an Embodied Conversational Agent (ECA) communicates better with users when interacting with non-verbal behavior, thus making the interaction more efficient and natural. The authors~\cite{nagy2021framework} proposed a model applied to conversational agents, including data-driven gesture generation models and end-to-end gesture synthesis models. They created a 3D conversational agent in Unity, using resources that convert text to speech and neural networks to generate speech-driven body movements. The user can interact by communicating through text or voice commands. The agent responds both in text and audio and performs body gestures.  

In the paper of Cassell et al.~\cite{cassell1994animated}, the authors present a system that automatically generates and animates conversations between different agents. These agents have speech, gestures, facial expressions, and intonation similar to humans in a synchronized manner, allowing for the creation of interactive dialogue animation. The developed system uses a face-to-face interaction model to generate the implemented behaviors, including speech intonation, hand gestures, gaze, and head nods. According to the authors~\cite{cassell1994animated}, the way that speech is performed and the intonation generated by the speaker are important in determining head movements, hand gestures, and facial expressions. Furthermore, Cassell et al.~\cite{cassell1994animated} argue that realistic autonomous agents cannot exist without verbal and non-verbal behavior.

Sajjadi et al.~\cite{sajjadi2019personality} analyzed how personality-based conversational agents affect social presence in a virtual reality simulation. In this research, the conversational agent Linda was developed. They synthesized a range of facial expressions, gestures, and postures for the agent to communicate non-verbally. The simulation used to evaluate the Embodied Conversational Agent (ECA) was an office environment where users were tasked with reviewing Linda's performance in the company. In each experiment, Linda could have one of three personalities: neutral, extroverted, and introverted. The study concluded that the extroverted personality led to greater behavioral engagement with the users than neutral and introverted personalities. According to the authors~\cite{sajjadi2019personality}, this occurred because non-verbal behaviors are more noticeable and assertive with an extroverted personality.

Cassell~\cite{cassell2000embodied} proposes the creation of a conversational agent called Rea. This agent looks human-like and behaves like a human during a dialogue. The agent utilizes gaze, facial expressions, body posture, and hand movements to engage in conversation. Rea is projected on a screen, and, to interact directly with her, the user must communicate using a microphone. According to the author~\cite{cassell2000embodied}, gestures and body posture are essential as they play a fundamental role in communication, such as initiating and ending a conversation, enabling real-time interaction. Therefore, Rea is an agent capable of producing verbal and non-verbal behaviors, enhancing the user's interaction during a dialogue.

He et al.~\cite{he2022evaluating} proposed a study in which users could interact directly with two ECAs, one with gestures and the other without gestures. According to the authors~\cite{he2022evaluating}, in many studies, participants are only asked to watch pre-recorded videos, which do not provide as relevant information about the users' perception during the interaction. In this model, users only participate as observers and do not interact directly with the conversational agents. According to the authors~\cite{he2022evaluating}, this model is not commonly applied in real-world scenarios, and participants perceive more details when interacting. The work aims to evaluate the effectiveness of gesture generation and determine if the movements generated by the model affect how users perceive the conversational agent during the interaction. Therefore, they developed a study that aims to assess the user experience during real-time interaction, which is considered an effective method for evaluating the performance of the ECA. According to the study, the results did not indicate significant differences. The ECA with gestures may attract more attention from participants for specific topics. However, some users reported excessive body movements, which may have distracted their attention, taking the focus away from the users, unlike the agent without gestures, which allowed participants to concentrate on what was being presented.

Sonlu et al.~\cite{sonlu2021conversational} developed a conversational agent that expresses personality through verbal and non-verbal behavior. According to the authors~\cite{sonlu2021conversational}, people evaluate the agent's personality based on speech and body movements and react to them as if they were human. They conducted a study with participants to assess their opinions regarding the agent. Although they did not carry out real-time interactions, they used videos to show to the users and evaluate their perceptions. According to the results, users considered speech as base on the naturalness of the movements and facial expressions. The study also suggests that the combination of dialogue, voice, and body movements contributes to the perception of the agent's personality.

The work of Knob et al.~\cite{knob2021arthur} proposes an Embodied Conversational Agent called Arthur. This agent communicates with the user and can detect facial emotions and recognize who is speaking. Arthur has an artificial memory responsible for storing and retrieving data about specific events based on the model of human memory. The authors~\cite{knob2021arthur} conducted a study to gather information about the impact of the agent on users. According to the study, Arthur exhibited the expected behavior, and the implemented memory model was satisfactory. Regarding the agent's appearance, it was mentioned that it could still be improved. In the previous paper, Arthur is portrayed as a 2D cartoon. 
%PA: essa parte me parece que fica ruim no artigo.
%The authors~\cite{knob2021arthur} made some improvements to make the agent's appearance more human-like.
%SO: sobre acima.. era sobre o Arthur 2D, e abaixo é sobre o new Arthur certo?
%PA: então, não sei... acho que era falando já do novo... O parágrafo abaixo me parece que fca melhor.

There are similar works in the literature, as most of the related works compare conversational agents with and without body movements, and the results obtained with users who interacted with those agents. The difference, in relation to ours, is that in related works, users interact only via video or real time, but not both: thus, they are not able to make a comparison. Another difference is the technology used. None of the works found in the literature used Unreal Engine to model their virtual agents, nor Metahuman.
In the present paper, we evaluated the impact of body animation on a new version of Arthur, which is more realistic.
%The study presented in this paper was essential for the development of this project, as it served as the basis for creating the body animations for a more realistic Arthur. 
Building upon the facial animations implemented, we decided to incorporate body movements into the agent, as described in Section~\ref{sec:metodologia}.

% Wolfert, Robinson, Belpaeme (2022) fizeram uma revisão de estudos que utilizam agentes de conversação incorporados. De acordo com Wolfert, Robinson, Belpaeme (2022), um componente importante na comunicação não verbal é a utilização de gestos, como o movimento das mãos, braços ou corpo, é por meio deles que a mensagem é enfatizada. Ainda de acordo com o artigo, os humanos estão mais dispostos a interagirem quando um ECA realiza gestos apropriados do que quando não utilizam gestos ou não corresponde com a comunicação verbal. Desta forma, Wolfert, Robinson, Belpaeme (2022) fizeram uma revisão com 22 estudos que utilizaram ECAs que possuem um corpo semelhante à forma humana que usa co-fala e gesticula na interação entre humano-agente. Na análise feita por eles, metade dos estudos utilizou avatares que mostravam apenas a parte superior do corpo, enquanto que a outra mostrou o corpo inteiro. O número de participantes variou e muitos não mencionaram estatísticas básicas sobre os participantes, sendo que poucos estudos relataram o método de avaliação detalhado. Em relação a qual metodologia é mais eficaz, nenhuma forma de avaliação foi considerada melhor do que a outra. Ainda de acordo com o artigo, é recomendado que utilizem métodos objetivos e subjetivos, sendo que as avaliações subjetivas devem ser com diversas populações e origens. A literatura não mostrou um uso consistente nas métricas de gestos. A maior parte dos casos utilizou como método de avaliação um questionário, para avaliar a qualidade dos gestos de co-fala em ECAs. 
%

\section{Proposed Model}
\label{sec:metodologia}

\begin{figure}[htp]
    \centering
    \includegraphics[width=8cm]{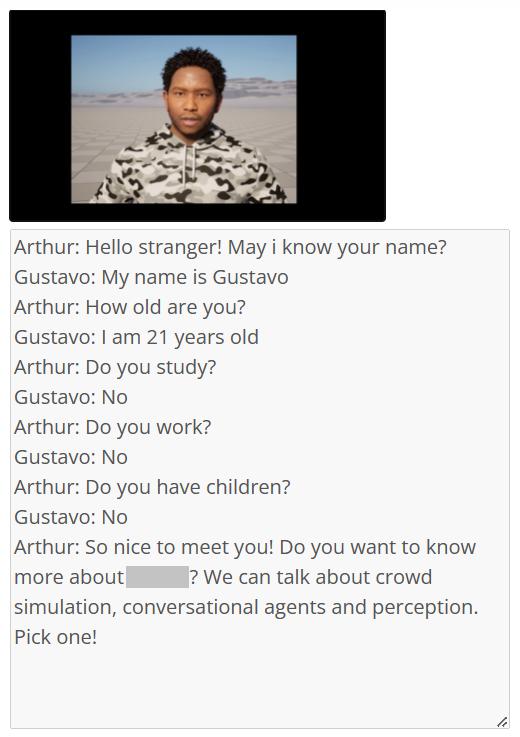}
    \caption{\small Example of interaction with the conversational agent. The name of the Research Lab was omitted for blind review.}
    \label{fig:interacao}
\end{figure}

This work aims to develop arms and hand movements for a conversational agent. This research was based on the agent developed by Knob et al.~\cite{knob2021arthur}, Arthur, as described in Section~\ref{sec:relatedwork}. The provided model only contained facial animations developed using the MetaHuman framework\footnote{\url{https://www.unrealengine.com/en-US/metahuman}}, developed by Unreal Engine.
%SO: made using Unreal??? tem que dizer algo aqui
%PA: adicionei essa info
From there, we developed the rest of the body movements. Figure~\ref{fig:interacao} illustrates an example of an interaction with Arthur. At the top, it shows the video animation and, below it, the chat through which the user can interact with the virtual agent.

This section is divided into two sections. Section~\ref{sec:geracao_anim} explains how the body animations for Arthur were generated, while Section~\ref{sec:questionario} describes the questionnaires that were used to evaluate the users' perception concerning the conversational agent.

% \begin{figure}[htp]
%     \centering
%     \includegraphics[width=8cm]{john.png}
%     \caption{\small Agente conversacional Arthur}
%     {\footnotesize{Fonte: \cite{knob2022modeling}}}
%     \label{fig:john}
% \end{figure}

%SO: essa figura abaixo tem que riscar o VHLAB por causa do blind. Explicar isso no caption da fig, "The name of the Research Lab was omitted for blind review"
%PA: feito

\subsection{Animations' Generation}
\label{sec:geracao_anim}

%PA: aqui, é sobre a Unreal ou sobre o Meta-Humans?
%SO: metahuman é da unreal não?
%PA: Sim, mas o MH tem sua própria "marca", por assim dizer. Se for falar no MH, acho melhor já falar algo como "MetaHuman, developed by Unreal Engine,..."
The MetaHuman framework, developed by Unreal Engine, is an advanced development environment that can be used to create virtual humans~\cite{origlia2022developing}. Since the implementation of facial animations was done on this platform, we chose to continue using it for the project's development. The animations were created based on the dialog developed by Knob et al.~\cite{knob2021arthur}. The initial project had facial animations synchronized with lip movement and speech. Building upon this project, we implemented our animations for Arthur's body.

%PA: De novo: o artigo não fala desse agente no python, então não são apenas as 8 animações
Arthur is a conversational agent that interacts through chat, audio, and predefined videos. In Knob et al.'s full model~\cite{knob2021arthur}, many possible dialogues and sentences are in the user conversation. To control which dialog the users should interact with, we restricted the possible dialogues to a subset containing only eight facial animation videos that cover specific topics with users. Therefore, we created body movements for these eight animations, summing up to a total of 16 animations. Because of the amount of participants in our experiment (more details in Section~\ref{sec:resultados}), we thought it would be better to not create more complex tasks with more dialogs (which would generate more facial animations to be evaluated). The goal was to have the same dialogue for all users, with animations that could include body and facial motion or only facial one.

\begin{figure}[htp]
    \centering
    \includegraphics[width=8cm]{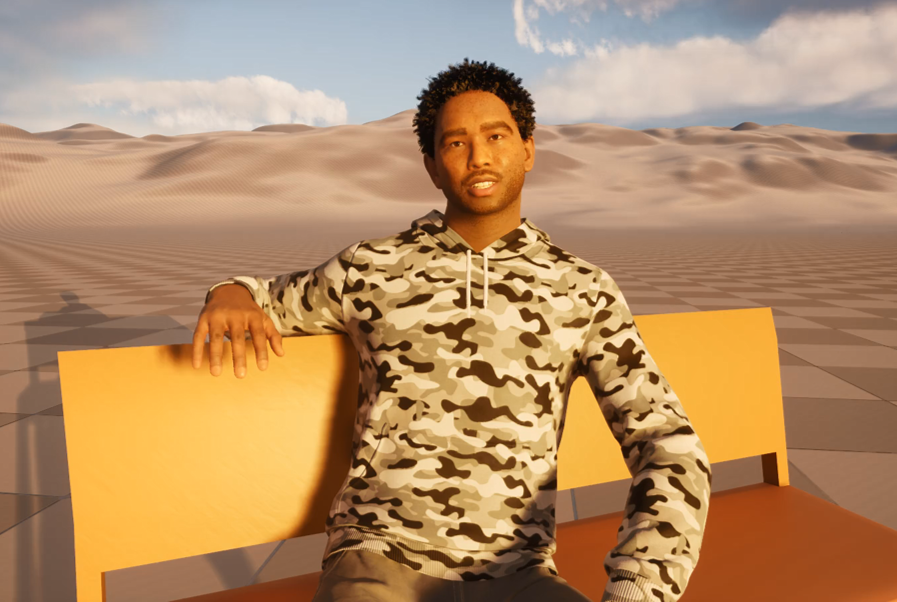}
    \caption{\small Illustration of a body animation designed for our Embodied Conversational Agent.}
    \label{fig:arthur}
\end{figure}

%We used a free animation available in Unreal, which only included upper body movements of the limbs. As a result, w
We decided to develop animations solely for the arms and hands without including leg movements. Figure~\ref{fig:arthur} shows Arthur with an upper body and facial animations, where he initiates a dialogue with the user. He is seated on a bench and performs gestures, moving the upper part of his body, including his hands, arms, and head, in addition to the facial motion.

\begin{figure}[htp]
    \centering
    \includegraphics[width=7.4cm]{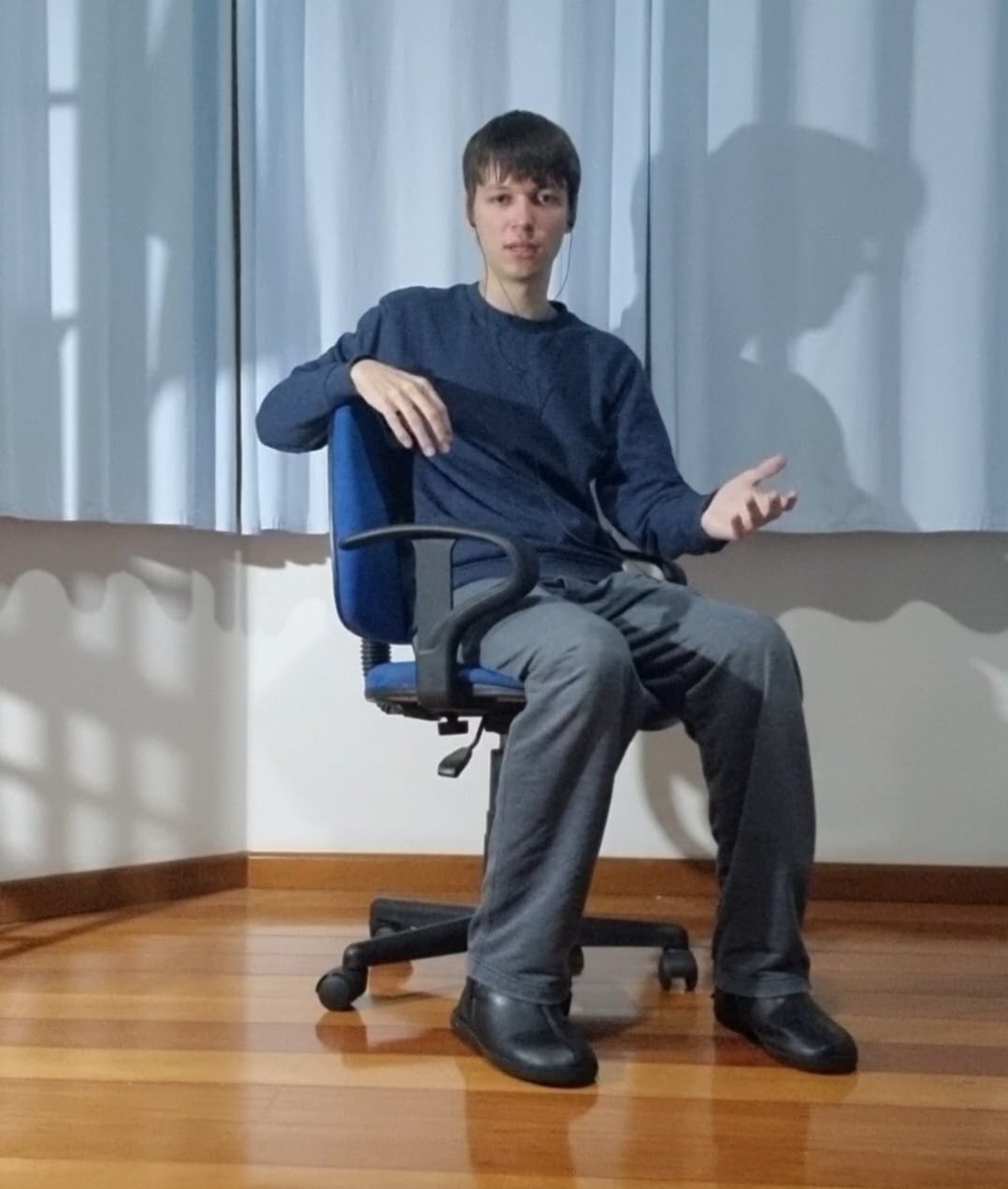}
    \caption{\small Image obtained from one of the videos recorded, performing the body movements for the animations.}
    \label{fig:movimentos}
\end{figure}

We first used one available upper body animation shown in Figure~\ref{fig:arthur} for a situation where the character is sitting. This free animation was integrated with the facial animation we had for a specific sentence in the conversation. In addition, we produce another video, with the same facial animation and scenario, but without the body animation, to be used in the comparisons.
%This animation was generated in order to analyze users' perceptions regarding animations with and without body movements. 
Subsequently, we implemented the remaining seven animations to complete the subset of the dialogue chosen from the work of Knob et al.~\cite{knob2021arthur}. Since there were no other body animations available in Unreal that were similar to the first one, % but different from the one that was used, 
we chose to generate our own animations. 
%In order to develop our own body animations, w
We used DeepMotion\footnote{\url{https://deepmotion.com/}}, which transforms videos into 3D animations by capturing full-body movements. Thus, we used this tool to generate body animations for the conversational agent. Firstly, we recorded our videos in which we performed arm and hand movements and later inserted the animations into Arthur. We always generated two versions: one with the facial and body animation, and the other only with facial animation, but in the same scenario. Figure~\ref{fig:movimentos} shows an image from one of the videos, demonstrating how the movements were filmed. Based on the agent's dialogue, we made body movements that were consistent, in the opinion of our research group, with the speech. We tried to do simple and generic movements, that should not impact the participants' perception. We recorded seven videos, generating non-verbal behaviors related to Arthur's dialogues selected, as mentioned before. %The other video had been implemented using the animation available in Unreal. 
After finalizing and generating the 3D animations in DeepMotion, the files were imported into Unreal and added to Arthur's movements. 

At the end of the implementation, we rendered all the animations to generate the videos of the conversational agent to be played during the  interaction with the users. In order to do so, we used Unreal, which has a plugin for video rendering. The platform generates a file containing multiple images in PNG format. Therefore, to combine the images with the audio, we used the FFmpeg tool\footnote{\url{https://www.ffmpeg.org/}}, which merged the images together with the audio, generating a video file.

\subsection{Questionnaire with the Users}
\label{sec:questionario}

To evaluate the users' perception regarding the presence or not of the body animations, three questionnaires were created. %and conducted for users to assess the conversational agent. 
%Therefore, three questionnaires were created: 
The first one evaluated the perception of users who do not interact with the ECA (just watched the videos), and two others where the users interacted directly with the ECA. In the non-interactive questionnaire, users compared two videos of the same agent, one without body animations and the other with body animations. In the other two questionnaires, users directly interacted with Arthur by chat. One questionnaire featured the agent with body animations, while the other had only facial animations. 

%This research had the limitation of using only a single website link for the interactions with the virtual agent. For this reason, we created two questionnaires. In one of them, participants answered questions regarding the agent without body animations. After completing the data collection for the non-interactive questionnaire, we updated the virtual agent, which used the same link for interaction, but the animations were changed to the version with body movements. Thus, the same link was made available for the conversational agent, but at different times.

%PA: esse termo é o mesmo que o Victor criou? Acho que temos que colocar as informações dele aqui, certo? Deixar comentado aqui embaixo já.
%the participants were presented with the consent form approved by the Ethics Committee of University of Pontifical Catholic University of Rio Grande do Sul, referring to the research project entitled ``Estudos e Avaliações da Percepção Humana em Personagens e Multidões Virtuais", number $46571721.6.0000.5336$.
The three questionnaires begin with the informed ethical and consent form, followed by demographic questions, and finally, questions related to the agent. The collected demographic information included gender, age, level of education, and familiarity with computer graphics. The participants were presented with the consent form approved by the Ethics Committee of University~\footnote{Number and name are omitted for blind review.}.

% sem interacao
In the first questionnaire (i.e., when the user did not interact with the virtual agent), after the demographic questions, the user is invited to watch two videos. In video 1, Arthur only speaks while applying facial animations, while in video 2, he has both facial and body animations. After watching the videos, the user is invited to answer the following questions:

\begin{itemize}
    \item Q1: Which video did you like the most?
    \item Q2: What is your level of satisfaction concerning the communication abilities of the virtual agent in Video 1?
    \item Q3: What is your level of satisfaction concerning the communication abilities of the virtual agent in Video 2?
    \item Q4: What is your level of satisfaction concerning the animations of the virtual agent in Video 1?
    \item Q5: What is your level of satisfaction concerning the animations of the virtual agent in Video 2?
\end{itemize}

For Q1, the possible answers are "Video 1" and "Video 2". For the remaining four questions, the answers are given following a Likert scale from 1 to 5, being 1 = "Very Dissatisfied", and 5 = "Very Satisfied".

% com interacao
In the second and third questionnaires (when the user interacts with the virtual agent), after the demographic questions, the user is invited to interact with Arthur. The interaction lasts for at least, one minute. However, if the user desires, he/she can keep interacting with the ECA as long as he/she wants. After the interaction, the users answer the following questions:

\begin{itemize}
 \item Q1: What is your level of satisfaction concerning the communication abilities of the conversational agent?
  \item Q2: What is your level of satisfaction concerning the interaction abilities of the conversational agent?
    \item Q3: What is your level of satisfaction concerning the animations of the conversational agent?
    \end{itemize}

%    \item Q1: Qual o seu nível de satisfação em relação a interação com o agente conversacional?
%    \item Q2: Qual o seu nível de satisfação em relação a comunicação do agente conversacional?
%    \item Q3: Qual o seu nível de satisfação em relação a animação do agente conversacional?
%\end{itemize}

%Due to the restriction of conducting only one questionnaire at a time for interactions, it was not possible to allow users to directly compare the interactions, as in question Q1 of the first questionnaire. Therefore, the same questions were used in both interactive questionnaires, so that the results could be compared. 
The answers to these three questions are Likert scales, ranging from 1 (Very Dissatisfied) to 5 (Very Satisfied). The next section presents the results obtained in this work.

\section{Results}
\label{sec:resultados}

This section presents the analysis and interpretation of the results obtained. % in a quantitative manner. 
Section~\ref{sec:análise} provides some qualitative examples of the interaction with the user, while Section~\ref{sec:analise_questionario} presents the results obtained from the application of the questionnaires. The evaluation will allow us to analyze users' perceptions regarding the conversational agent, as well as their opinions about Arthur, with and without body movements. It will also be possible to verify if there were differences in the results when participants watched the videos and when they were interacting directly. Finally, evaluating the user experience is essential to analyze the proposed model and understand the reasons behind the obtained results. This will allow us to explore alternative solutions and identify potential improvements for future work.

\subsection{Qualitative Example of Interaction}
\label{sec:análise}

\begin{figure}[htp]
    \centering
    \includegraphics[width=8cm]{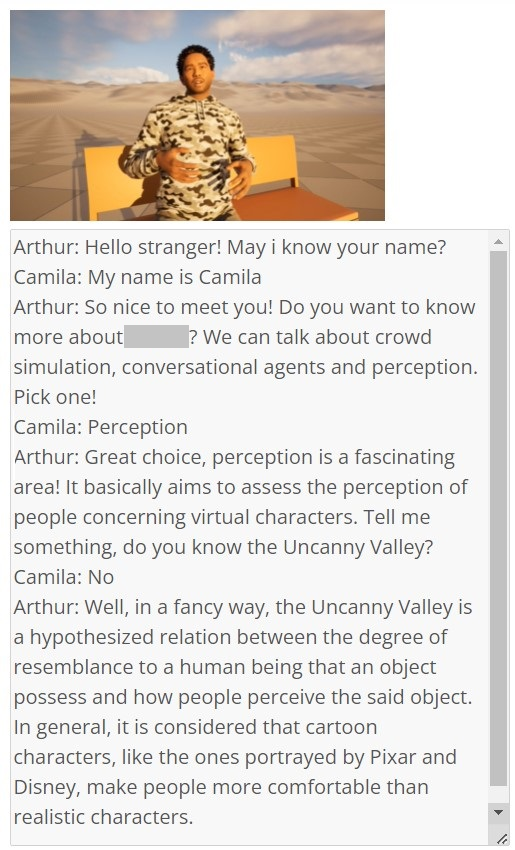}
    \caption{\small An example of an interaction with Arthur portraying body movements as a non-verbal behavior. The name of the Research Lab was omitted for blind review.}
    \label{fig:exemploInteracao}
\end{figure}

\begin{figure}[htp]
    \centering
    \includegraphics[width=8cm]{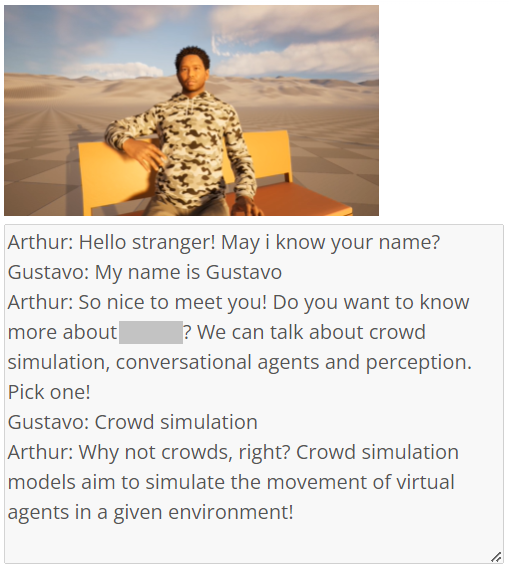}
    \caption{\small An example of an interaction with Arthur portraying body movements as a non-verbal behavior. The name of the Research Lab was omitted for blind review.}
    \label{fig:exemploInteracao2}
\end{figure}

\begin{figure}[htp]
    \centering
    \includegraphics[width=8cm]{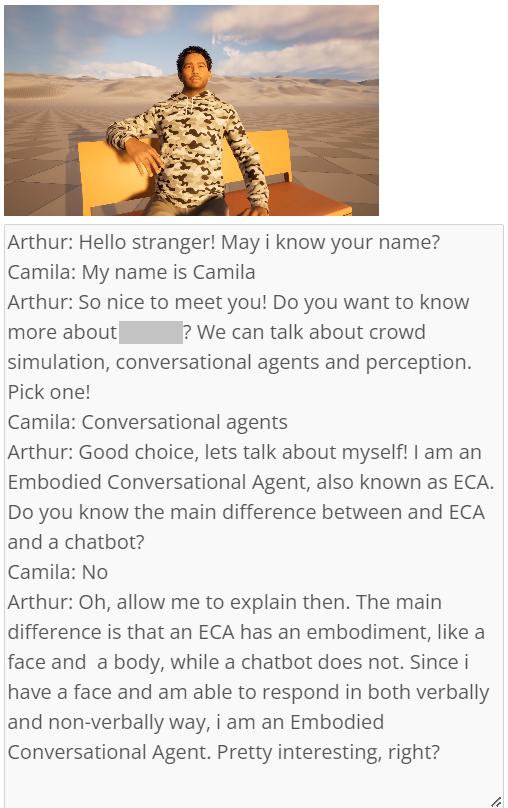}
    \caption{\small An example of an interaction with Arthur portraying body movements as a non-verbal behavior. The name of the Research Lab was omitted for blind review.}
    \label{fig:exemploInteracao3}
\end{figure}

As described in Section~\ref{sec:geracao_anim}, we generated body animations for Arthur. As a result, the agent is able to move his arms, hands, and head, as well as perform facial expressions when interacting with the user. Figures~\ref{fig:exemploInteracao}, \ref{fig:exemploInteracao2} and \ref{fig:exemploInteracao3} show examples of interactions with Arthur after the implementation of the upper limbs, as described in Section~\ref{sec:geracao_anim}. The bottom part displays the chat interface, where the user can communicate with him.
For participants who interacted with the conversational agent in real time, a link was provided for the interaction. There was no need to install anything on their devices, they just had to be connected to the internet. We recommended using a personal computer or notebook to facilitate viewing the conversational agent, as the cell phone screen is smaller. We suggested that the interaction should last for, at least, one minute, but there was no maximum time limit. The collected data were obtained through the answers given in the questionnaires by the users.
%SO: seria legal nesta seção colocar mais umas duas figuras no mesmo estilo.
%PA: não fui eu, mas me parece que feito =D

%PA: aqui embaixo é citado o arthur novamente, mas o modulo implementado para tal não eh citado naquele trabalho.
During the interaction, the agent invites the user to choose three topics to discuss, such as crowd simulation, conversational agents, and perception. The user responds through the chat interface, and the agent replies using audio, text, and video animation, as illustrated in Figures~\ref{fig:exemploInteracao}, \ref{fig:exemploInteracao2} and \ref{fig:exemploInteracao3}. The video is chosen based on the sentence that the agent selects as a response, which is determined by the module implemented by Knob et al.~\cite{knob2021arthur}. If the user communicates in a way that does not include one of the proposed topics by Arthur, in the chosen subset to provide a controlled interaction, the video ends, and the agent informs that it was not possible to understand what the user meant. However, the participant can restart the conversation and continue interacting with the agent. 

%SO: É importante riscar o VHLab da figura
%PA: feito

\subsection{Questionnaires's Analysis}
\label{sec:analise_questionario}

%PA: Por que 30? Soraia: please see the works proposed by Prof. Rachel Macdonell who works on perception area about virtual humans.
We obtained 30 answers for the first questionnaire, with 17 male participants and 13 female participants ranging in age from 16 to 83 years old. Regarding the analyzed videos (Q1) containing the agent with or without body animation, 96.7\% of participants preferred video 2, in which Arthur performs body movements. Only one user opted for video 1. There is a contrast among the participants regarding familiarity with computer graphics, with 10 indicating high familiarity and 8 indicating low familiarity. The remaining participants responded on the Likert scale ranging from 2, 3, and 4. 

Table~\ref{tab:tab1} displays the responses obtained for each of the remaining questions. The lines labeled Q2, Q3, Q4, and Q5 correspond to the questions asked to the users, as explained in Section~\ref{sec:questionario}. The highest values for each question are marked in bold.

\begin{table}[htb]
    \centering
    \begin{tabular}{c|c c c c c| c c c}
        {} & 1 & 2 & 3 & 4 & 5 & Avg & Median & $\Sigma$\\
        \hline
        Q2 & 7 & 8 & \bfseries 10 & 4 & 1  & 2.46 & 2.50 & 1.10\\
        Q3 & 2 & 2 & 5 & \bfseries 14 & 7 & \textbf{3.73} & \textbf{4.00} & 1.11\\
        Q4 & \bfseries11 & \bfseries11 & 4 & 4 & 0 & 2.03 & 2.00 & 1.03\\
        Q5 & 2 & 1 & 8 & \bfseries 11 & 8 & \textbf{3.73} & \textbf{4.00} & 1.11\\
    \end{tabular}
    \caption{Answers obtained from the first questionnaire (only videos). Q2 and Q4 are related to the videos without body movements, while Q3 and Q5 are related to the videos with body movements. Header values (i.e., 1, 2, 3, 4, and 5) represent the Likert Scale values. The highest values for each question (i.e., Q2, Q3, Q4, and Q5) are marked in bold. The three columns on the right present each question's average, median, and standard deviation values.}
    \label{tab:tab1}
\end{table}

In Table~\ref{tab:tab1}, Q2 and Q4 assess the video in which the agent does not have body animation, while Q3 and Q5 evaluated the agent with body animation. %Q4 assesses the level of satisfaction regarding the animation of the conversational agent in Video 1 (without body animation). 
The average, median, and standard deviation values are presented in the table. We can see that the higher values were obtained for Q3 and Q5 (indeed the same value). We conclude that users can perceive and are impacted by the body animation of the conversational agent.
We performed a two-tailed Mann-Whitney test in order to compare these two groups in the first questionnaire (i.e., with/without body movements), which resulted in a p-value of 0.00001, which indicates a significant difference between the two groups.

%The median of participants' responses in Q4 was 2, indicating that the animation was considered unsatisfactory. %, as no respondents selected the option of very satisfied. However, for the animation in Video 2, the median was 4, indicating it was considered satisfactory, with 8 users choosing 5 on the Likert scale. Analyzing the responses regarding participants' opinions in the first questionnaire, it is observed that the animations in video 2 left users more satisfied. Thus, it can be noted that the body movement made users feel more comfortable when interacting with Arthur.

In the second questionnaire, participants interacted with Arthur who %interacts with the user but 
does not perform body movements. In this questionnaire, we obtained only 15 responses, with ten female and five male participants. %This difference in the number of participants compared to the first questionnaire occurred due to the restriction of the link, as only one person could interact at a time. Therefore, access to the link for interaction with the agent was restricted to prevent simultaneous access. As a result, a smaller number of users agreed to participate in this study. 
The participants' age ranged from 16 to 53 years old. Regarding familiarity with computer graphics, five participants reported having a high level of familiarity, while four reported having a low level. The remaining participants selected responses ranging from 2, 3, to 4 on the Likert scale. Table~\ref{tab:tab2} displays the answers obtained for each question. %The values in columns Q1, Q2, and Q3 correspond to the questions posed to the users, as explained in Section~\ref{sec:questionario}. 
The highest values for each question are marked in bold. In addition, in the last three columns, we included the average, median, and standard deviation values for each question.

%This questionnaire will be analyzed next, alongside the presentation of the final questionnaire.

\begin{table}[htb]
    \centering
    \begin{tabular}{c|c c c c c|c c c}
        {} & 1 & 2 & 3 & 4 & 5 & Avg & Median & $\Sigma$\\
        \hline
        Q1 & 0 & 4 & \bfseries 8 & 1 & 2 & 3.06 & 3.00 & 0.96\\
        Q2 & 0 & \bfseries 6 & 4 & 4 & 1 & 3.00 & 3.00 & 1.00\\
        Q3 & 0 & \bfseries 6 & 3 & 2 & 4 & 3.26 & 3.00 & 1.27\\
    \end{tabular}
    \caption{Answers obtained in the second questionnaire (interactive agent without body animation). Header values (i.e., 1, 2, 3, 4, and 5) represent the Likert Scale values. The highest values for each question (i.e., Q1, Q2, and Q3) are marked in bold. The three columns on the right present each question's average, median, and standard deviation values.}
    \label{tab:tab2}
\end{table}

Finally, the third and final questionnaire, in which Arthur interacts with the participant by performing body movements, obtained 15 responses, with 9 participants female and 6 male, ranging in age from 16 to 44 years old. Regarding familiarity with computer graphics, 9 participants reported having a high level of familiarity, while only one user stated having a low level. The remaining respondents selected responses ranging from 3 to 4 on the Likert scale. In this questionnaire, no one answered 2 regarding familiarity with computer graphics. The highest values for each question are marked in bold. Table~\ref{tab:tab3} displays the responses obtained from the questions posed in the questionnaire. The column values Q1, Q2, and Q3 correspond to the questions asked to the users, as explained in Section~\ref{sec:questionario}. In addition, the last three column shows the average, median, and standard deviation values for each question.

\begin{table}[htb]
    \centering
    \begin{tabular}{c|c c c c c|c c c}
        {} & 1 & 2 & 3 & 4 & 5 & Avg & Median & $\Sigma$ \\
        \hline
        Q1 & 0 & 2 & 5 & \bfseries 6 & 2 & 3.53 & 4.00 & 0.91\\
        Q2 & 0 & 4 & 2 & 4 & \bfseries 5 & 3.66 & 4.00 & 1.23\\
        Q3 & 0 & 0 & 4 & \bfseries 7 & 4 & 4.00 & 4.00 & 0.75\\
    \end{tabular}
    \caption{Answers obtained in the third questionnaire (interactive agent with body animation). Header values (i.e., 1, 2, 3, 4, and 5) represent the Likert Scale values. The highest values for each question (i.e., Q1, Q2, and Q3) are marked in bold. The three columns on the right present each question's average, median, and standard deviation values.}
    \label{tab:tab3}
\end{table}

%In all three questions of the second questionnaire, the median of the responses was 3, indicating that users were neither satisfied nor dissatisfied with the interaction, communication, and animations of Arthur. For the third questionnaire, all three questions increased to a median of 4, suggesting that the addition of body animations made the interaction, communication, and animations more satisfactory for the users. 
%PA: nessa ultima frase, por que? Isso seria um resultado geral dos tres questionarios? Explicitar.

Comparing Tables~\ref{tab:tab2} and~\ref{tab:tab3}, we can observe that average and median values are higher for all questions when the interactive conversational agent applies body animation. In both questionnaires, Q3 obtained higher evaluation values, indicating that participants considered Arthur's animation better than his ability to communicate or interact. 
Although it can indicate that body movements enhance user perception, a significant difference was not observed when performing a two-tailed Mann-Whitney test, resulting in a p-value of 0.105.
%Thus, it can be observed that body movements enhance user perception, regardless of whether it is a video or an interaction. 

Analyzing the differences in user perception during interaction with Arthur or just watching his videos, it is noticed that the animations were considered more satisfactory for the conversational agent with body movements, regardless of whether the user interacted via video or directly.
We performed a two-tailed Mann-Whitney test in order to compare these two groups (i.e. with/without body movements), which resulted in a p-value of 0.00001, which indicates a significant difference between the two groups.
%PA: não consigo calcular esse, pela diferença de amostragem
On the other hand, if Arthur does not perform body movements, there is a difference in the perception of users who watched the video and those who interacted: the animations were more satisfactory for those who interacted directly. One possible explanation is that the chat feature prevented users from constantly looking at the agent, and therefore they were not bothered by the lack of body movements. Another possibility is that users who engaged in the interaction could not compare Arthur with and without body movements because they answered different questionnaires. 
%PA: animações sem movimentos? Não entendi. Ainda, Soraia: o Victor não tinha alguma literatura sobre isso, dizendo que interações são "melhores" que vídeos? 
%SO: Não sei, temos que ver com ele.
%PA: Pelo que conversei com ele, não tem nada nesse sentido.
Therefore, participants who interacted with the agent considered the animations without body movements more satisfactory than the participants who watched the video, as the latter had both animations as a basis for comparison of quality.
Finally, we also compared all answers from questionnaires 2 and 3, comparing such answers to discover if there was any difference in what participants perceived between interactions without body movements (questionnaire 2) and with body movements (questionnaire 3). We performed a two-tailed Mann-
Whitney test in order to compare these two groups, which resulted
in a p-value of 0.001, which indicates a significant difference
between the two groups.

\section{Final Considerations}
\label{sec:conclusao}

This paper analyzed users' perceptions regarding the nonverbal communication of a conversational agent. To do so, body animations were generated for Arthur~\cite{knob2021arthur}, which were evaluated by users through the application of three questionnaires. In one questionnaire, participants watched videos of the virtual agent, while in the other two questionnaires, they interacted directly with it. Through the analysis of the obtained results, it can be concluded that the non-verbal behavior of the conversational agent (i.e., body movements) brings more satisfaction to the users with respect to its communication and animation abilities. %made the interactions more comfortable. 
%It happened when the users watched the videos, but also when participants interacted with the agents.
Indeed, the developed animations for Arthur, performing hand and arm movements, were more satisfactory for users, %when compared to the absence of movements, 
regardless of whether the user interacted directly or via video. Furthermore, all the participants who interacted with Arthur and those who watched the video perceived an improvement in the agent's communication when it performed body movements. 

The obtained results demonstrate that including hand and arm movements altered users' perception of Arthur, as the level of satisfaction increased compared to the agent without movements. Therefore, the inclusion of body movements enhanced the user experience, making them feel more comfortable when interacting via video or directly in real-time.

%This work was limited in having only one interaction being performed with the virtual agent at the same time. 
This work has some limitations. Firstly, we want to proceed with the research and have more participants, which will allow us to analyze statistically our results. Therefore, as future work, it is worth considering the possibility of having two (or more) versions of the conversational agent, e.g., a woman, and in different scenarios. %In this way, the same user can interact with both versions of Arthur to compare which interaction they preferred, as was done in the questionnaire with the videos. 
Finally, another possibility is to address other aspects of the interaction with the conversational agent, such as the impact of leg movement on user perception.

\bibliographystyle{ACM-Reference-Format}
\bibliography{sample-base}

%%% -*-BibTeX-*-
%%% Do NOT edit. File created by BibTeX with style
%%% ACM-Reference-Format-Journals [18-Jan-2012].

\begin{thebibliography}{15}

%%% ====================================================================
%%% NOTE TO THE USER: you can override these defaults by providing
%%% customized versions of any of these macros before the \bibliography
%%% command.  Each of them MUST provide its own final punctuation,
%%% except for \shownote{}, \showDOI{}, and \showURL{}.  The latter two
%%% do not use final punctuation, in order to avoid confusing it with
%%% the Web address.
%%%
%%% To suppress output of a particular field, define its macro to expand
%%% to an empty string, or better, \unskip, like this:
%%%
%%% \newcommand{\showDOI}[1]{\unskip}   % LaTeX syntax
%%%
%%% \def \showDOI #1{\unskip}           % plain TeX syntax
%%%
%%% ====================================================================

\ifx \showCODEN    \undefined \def \showCODEN     #1{\unskip}     \fi
\ifx \showDOI      \undefined \def \showDOI       #1{#1}\fi
\ifx \showISBNx    \undefined \def \showISBNx     #1{\unskip}     \fi
\ifx \showISBNxiii \undefined \def \showISBNxiii  #1{\unskip}     \fi
\ifx \showISSN     \undefined \def \showISSN      #1{\unskip}     \fi
\ifx \showLCCN     \undefined \def \showLCCN      #1{\unskip}     \fi
\ifx \shownote     \undefined \def \shownote      #1{#1}          \fi
\ifx \showarticletitle \undefined \def \showarticletitle #1{#1}   \fi
\ifx \showURL      \undefined \def \showURL       {\relax}        \fi
% The following commands are used for tagged output and should be
% invisible to TeX
\providecommand\bibfield[2]{#2}
\providecommand\bibinfo[2]{#2}
\providecommand\natexlab[1]{#1}
\providecommand\showeprint[2][]{arXiv:#2}

\bibitem[Biancardi et~al\mbox{.}(2017)]%
        {biancardi2017could}
\bibfield{author}{\bibinfo{person}{Beatrice Biancardi}, \bibinfo{person}{Angelo Cafaro}, {and} \bibinfo{person}{Catherine Pelachaud}.} \bibinfo{year}{2017}\natexlab{}.
\newblock \showarticletitle{Could a virtual agent be warm and competent? investigating user's impressions of agent's non-verbal behaviours}. In \bibinfo{booktitle}{\emph{Proceedings of the 1st acm sigchi international workshop on investigating social interactions with artificial agents}}. \bibinfo{pages}{22--24}.
\newblock


\bibitem[Breitfuss et~al\mbox{.}(2007)]%
        {breitfuss2007automated}
\bibfield{author}{\bibinfo{person}{Werner Breitfuss}, \bibinfo{person}{Helmut Prendinger}, {and} \bibinfo{person}{Mitsuru Ishizuka}.} \bibinfo{year}{2007}\natexlab{}.
\newblock \showarticletitle{Automated generation of non-verbal behavior for virtual embodied characters}. In \bibinfo{booktitle}{\emph{Proceedings of the 9th international conference on Multimodal interfaces}}. \bibinfo{pages}{319--322}.
\newblock


\bibitem[Cassell(2000)]%
        {cassell2000embodied}
\bibfield{author}{\bibinfo{person}{Justine Cassell}.} \bibinfo{year}{2000}\natexlab{}.
\newblock \showarticletitle{Embodied conversational interface agents}.
\newblock \bibinfo{journal}{\emph{Commun. ACM}} \bibinfo{volume}{43}, \bibinfo{number}{4} (\bibinfo{year}{2000}), \bibinfo{pages}{70--78}.
\newblock


\bibitem[Cassell et~al\mbox{.}(1994)]%
        {cassell1994animated}
\bibfield{author}{\bibinfo{person}{Justine Cassell}, \bibinfo{person}{Catherine Pelachaud}, \bibinfo{person}{Norman Badler}, \bibinfo{person}{Mark Steedman}, \bibinfo{person}{Brett Achorn}, \bibinfo{person}{Tripp Becket}, \bibinfo{person}{Brett Douville}, \bibinfo{person}{Scott Prevost}, {and} \bibinfo{person}{Matthew Stone}.} \bibinfo{year}{1994}\natexlab{}.
\newblock \showarticletitle{Animated conversation: rule-based generation of facial expression, gesture \& spoken intonation for multiple conversational agents}.
\newblock  (\bibinfo{year}{1994}), \bibinfo{pages}{413--420}.
\newblock


\bibitem[He et~al\mbox{.}(2022)]%
        {he2022evaluating}
\bibfield{author}{\bibinfo{person}{Yuan He}, \bibinfo{person}{Andr{\'e} Pereira}, {and} \bibinfo{person}{Taras Kucherenko}.} \bibinfo{year}{2022}\natexlab{}.
\newblock \showarticletitle{Evaluating data-driven co-speech gestures of embodied conversational agents through real-time interaction}. In \bibinfo{booktitle}{\emph{Proceedings of the 22nd ACM International Conference on Intelligent Virtual Agents}}. \bibinfo{pages}{1--8}.
\newblock


\bibitem[Knob et~al\mbox{.}(2021)]%
        {knob2021arthur}
\bibfield{author}{\bibinfo{person}{Paulo Knob}, \bibinfo{person}{Willian~S Dias}, \bibinfo{person}{Natanael Kuniechick}, \bibinfo{person}{Joao Moraes}, {and} \bibinfo{person}{Soraia~Raupp Musse}.} \bibinfo{year}{2021}\natexlab{}.
\newblock \showarticletitle{Arthur: a new ECA that uses Memory to improve Communication}. In \bibinfo{booktitle}{\emph{2021 IEEE 15th International Conference on Semantic Computing (ICSC)}}. IEEE, \bibinfo{pages}{163--170}.
\newblock


\bibitem[Kucherenko et~al\mbox{.}(2021)]%
        {kucherenko2021genea}
\bibfield{author}{\bibinfo{person}{Taras Kucherenko}, \bibinfo{person}{Patrik Jonell}, \bibinfo{person}{Youngwoo Yoon}, \bibinfo{person}{Pieter Wolfert}, \bibinfo{person}{Zerrin Yumak}, {and} \bibinfo{person}{Gustav Henter}.} \bibinfo{year}{2021}\natexlab{}.
\newblock \showarticletitle{GENEA Workshop 2021: The 2nd Workshop on Generation and Evaluation of Non-verbal Behaviour for Embodied Agents}. In \bibinfo{booktitle}{\emph{Proceedings of the 2021 International Conference on Multimodal Interaction}}. \bibinfo{pages}{872--873}.
\newblock


\bibitem[Nagy et~al\mbox{.}(2021)]%
        {nagy2021framework}
\bibfield{author}{\bibinfo{person}{Rajmund Nagy}, \bibinfo{person}{Taras Kucherenko}, \bibinfo{person}{Birger Moell}, \bibinfo{person}{Andr{\'e} Pereira}, \bibinfo{person}{Hedvig Kjellstr{\"o}m}, {and} \bibinfo{person}{Ulysses Bernardet}.} \bibinfo{year}{2021}\natexlab{}.
\newblock \showarticletitle{A framework for integrating gesture generation models into interactive conversational agents}.
\newblock \bibinfo{journal}{\emph{arXiv preprint arXiv:2102.12302}} (\bibinfo{year}{2021}).
\newblock


\bibitem[Origlia et~al\mbox{.}(2022)]%
        {origlia2022developing}
\bibfield{author}{\bibinfo{person}{Antonio Origlia}, \bibinfo{person}{Martina Di~Bratto}, \bibinfo{person}{Maria Di~Maro}, {and} \bibinfo{person}{Sabrina Mennella}.} \bibinfo{year}{2022}\natexlab{}.
\newblock \showarticletitle{Developing Embodied Conversational Agents in the Unreal Engine: The FANTASIA Plugin}. In \bibinfo{booktitle}{\emph{Proceedings of the 30th ACM International Conference on Multimedia}}. \bibinfo{pages}{6950--6951}.
\newblock


\bibitem[Pelachaud(2005)]%
        {pelachaud2005multimodal}
\bibfield{author}{\bibinfo{person}{Catherine Pelachaud}.} \bibinfo{year}{2005}\natexlab{}.
\newblock \showarticletitle{Multimodal expressive embodied conversational agents}. In \bibinfo{booktitle}{\emph{Proceedings of the 13th annual ACM international conference on Multimedia}}. \bibinfo{pages}{683--689}.
\newblock


\bibitem[Queiroz et~al\mbox{.}(2008)]%
        {Queiroz2008}
\bibfield{author}{\bibinfo{person}{Rossana~B. Queiroz}, \bibinfo{person}{Leandro~M. Barros}, {and} \bibinfo{person}{Soraia~R. Musse}.} \bibinfo{year}{2008}\natexlab{}.
\newblock \showarticletitle{Providing Expressive Gaze to Virtual Animated Characters in Interactive Applications}.
\newblock \bibinfo{journal}{\emph{Comput. Entertain.}} \bibinfo{volume}{6}, \bibinfo{number}{3}, Article \bibinfo{articleno}{41} (\bibinfo{date}{nov} \bibinfo{year}{2008}), \bibinfo{numpages}{23}~pages.
\newblock
\urldef\tempurl%
\url{https://doi.org/10.1145/1394021.1394034}
\showDOI{\tempurl}


\bibitem[Queiroz et~al\mbox{.}(2010)]%
        {Queiroz2010}
\bibfield{author}{\bibinfo{person}{Rossana~B. Queiroz}, \bibinfo{person}{Marcelo Cohen}, {and} \bibinfo{person}{Soraia~R. Musse}.} \bibinfo{year}{2010}\natexlab{}.
\newblock \showarticletitle{An Extensible Framework for Interactive Facial Animation with Facial Expressions, Lip Synchronization and Eye Behavior}.
\newblock \bibinfo{journal}{\emph{Comput. Entertain.}} \bibinfo{volume}{7}, \bibinfo{number}{4}, Article \bibinfo{articleno}{58} (\bibinfo{date}{jan} \bibinfo{year}{2010}), \bibinfo{numpages}{20}~pages.
\newblock
\urldef\tempurl%
\url{https://doi.org/10.1145/1658866.1658877}
\showDOI{\tempurl}


\bibitem[Sajjadi et~al\mbox{.}(2019)]%
        {sajjadi2019personality}
\bibfield{author}{\bibinfo{person}{Pejman Sajjadi}, \bibinfo{person}{Laura Hoffmann}, \bibinfo{person}{Philipp Cimiano}, {and} \bibinfo{person}{Stefan Kopp}.} \bibinfo{year}{2019}\natexlab{}.
\newblock \showarticletitle{A personality-based emotional model for embodied conversational agents: Effects on perceived social presence and game experience of users}.
\newblock \bibinfo{journal}{\emph{Entertainment Computing}}  \bibinfo{volume}{32} (\bibinfo{year}{2019}), \bibinfo{pages}{100313}.
\newblock


\bibitem[Sonlu et~al\mbox{.}(2021)]%
        {sonlu2021conversational}
\bibfield{author}{\bibinfo{person}{Sinan Sonlu}, \bibinfo{person}{U{\u{g}}ur G{\"u}d{\"u}kbay}, {and} \bibinfo{person}{Funda Durupinar}.} \bibinfo{year}{2021}\natexlab{}.
\newblock \showarticletitle{A conversational agent framework with multi-modal personality expression}.
\newblock \bibinfo{journal}{\emph{ACM Transactions on Graphics (TOG)}} \bibinfo{volume}{40}, \bibinfo{number}{1} (\bibinfo{year}{2021}), \bibinfo{pages}{1--16}.
\newblock


\bibitem[Xiao et~al\mbox{.}(2007)]%
        {xiao2007role}
\bibfield{author}{\bibinfo{person}{Jun Xiao}, \bibinfo{person}{John Stasko}, {and} \bibinfo{person}{Richard Catrambone}.} \bibinfo{year}{2007}\natexlab{}.
\newblock \showarticletitle{The role of choice and customization on users' interaction with embodied conversational agents: effects on perception and performance}. In \bibinfo{booktitle}{\emph{Proceedings of the SIGCHI conference on Human factors in computing systems}}. \bibinfo{pages}{1293--1302}.
\newblock


\end{thebibliography}

%%
%% If your work has an appendix, this is the place to put it.
% \appendix

% \section{Research Methods}

% \subsection{Part One}

% Lorem ipsum dolor sit amet, consectetur adipiscing elit. Morbi
% malesuada, quam in pulvinar varius, metus nunc fermentum urna, id
% sollicitudin purus odio sit amet enim. Aliquam ullamcorper eu ipsum
% vel mollis. Curabitur quis dictum nisl. Phasellus vel semper risus, et
% lacinia dolor. Integer ultricies commodo sem nec semper.

% \subsection{Part Two}

% Etiam commodo feugiat nisl pulvinar pellentesque. Etiam auctor sodales
% ligula, non varius nibh pulvinar semper. Suspendisse nec lectus non
% ipsum convallis congue hendrerit vitae sapien. Donec at laoreet
% eros. Vivamus non purus placerat, scelerisque diam eu, cursus
% ante. Etiam aliquam tortor auctor efficitur mattis.

% \section{Online Resources}

% Nam id fermentum dui. Suspendisse sagittis tortor a nulla mollis, in
% pulvinar ex pretium. Sed interdum orci quis metus euismod, et sagittis
% enim maximus. Vestibulum gravida massa ut felis suscipit
% congue. Quisque mattis elit a risus ultrices commodo venenatis eget
% dui. Etiam sagittis eleifend elementum.

% Nam interdum magna at lectus dignissim, ac dignissim lorem
% rhoncus. Maecenas eu arcu ac neque placerat aliquam. Nunc pulvinar
% massa et mattis lacinia.

\end{document}